\documentclass[%
 reprint,
superscriptaddress,
 amsmath,amssymb,
prl,
]{revtex4-1}

\usepackage{graphicx}% Include figure files
\usepackage{dcolumn}% Align table columns on decimal point
\usepackage{bm}% bold math
\usepackage{lipsum}

\usepackage[caption=false]{subfig}
\begin{document}

%\preprint{APS/123-QED}

% %%%%%%%%%%%%%%%%%%%%%%%%%%%%%% 75 character limit (including spaces)
%\title{Hamiltonian neural networks learn chaos and order}
\title{Physics enhanced neural networks predict order and chaos}

%\title{From islands of order to a sea of chaos: \\A Physics Savvy Neural Network Learns Nonlinear Dynamics}

% \title{From islands of order to a sea of chaos: \\Combining fundamental physics with machine learning to understand complex systems}

\author{Anshul Choudhary}
\affiliation{Nonlinear Artificial Intelligence Laboratory, Physics Department, North Carolina State University, Raleigh, NC 27607, USA }

\author{John F. Lindner*}
\affiliation{Nonlinear Artificial Intelligence Laboratory, Physics Department, North Carolina State University, Raleigh, NC 27607, USA }
\affiliation{Physics Department, The College of Wooster, Wooster, OH 44691, USA}

\author{Elliott G. Holliday}
\affiliation{Nonlinear Artificial Intelligence Laboratory, Physics Department, North Carolina State University, Raleigh, NC 27607, USA }

\author{Scott T. Miller}
\affiliation{Nonlinear Artificial Intelligence Laboratory, Physics Department, North Carolina State University, Raleigh, NC 27607, USA }

\author{Sudeshna Sinha}
\affiliation{Nonlinear Artificial Intelligence Laboratory, Physics Department, North Carolina State University, Raleigh, NC 27607, USA }
\affiliation{Indian Institute of Science Education and Research Mohali, Knowledge City, SAS Nagar, Sector 81, Manauli PO 140 306, Punjab,
India}

\author{William L. Ditto}
\affiliation{Nonlinear Artificial Intelligence Laboratory, Physics Department, North Carolina State University, Raleigh, NC 27607, USA }

\date{\today}

\begin{abstract}
Conventional artificial neural networks are powerful tools in science and industry, but they can fail when applied to nonlinear systems where order and chaos coexist. We use neural networks that incorporate the structures and symmetries of Hamiltonian dynamics to predict phase space trajectories even as nonlinear systems transition from order to chaos. We demonstrate Hamiltonian neural networks on the canonical H\'enon-Heiles system, which models diverse dynamics from astrophysics to chemistry. The power of the technique and the ubiquity of chaos suggest widespread utility.
\end{abstract}

\maketitle % generate title, including abstract

% %%%%%%%%%%%%%%%%%%%%%%%%%%%%%%
%\section{Introduction}
\textit{Introduction}.---Newton wrote, ``My brain never hurt more than in my studies of the Moon (and Earth and Sun)"~\cite{NewtonQuote}, thus anticipating that the seemingly simple three-body problem was intrinsically intractable. Nonetheless, Hamilton remarkably re-imagined Newton's laws as an incompressible  energy conserving flow in phase space~\cite{Hamilton}, and this formalism further highlighted the fundamental difference between integrable and non-integrable systems~\cite{Poincare}, heralding the revolutionary concept of classical chaos~\cite{Lorenz,Gleick}.

Today, artificial neural networks are popular tools in industry and academia~\cite{Haykin2008}, especially for classification and regression problems, and are beginning to elucidate nonlinear dynamics~\cite{Lusch2018} and fundamental physics~\cite{SciNet,AIFeynman,AIPhysicist}. Recent neural networks outperform traditional techniques in symbolic integration~\cite{anonymous2020deep} and numerical integration~\cite{breen2019newton} and outperform humans in strategy games like chess and Go~\cite{AlphaZero}. But neural networks have a blind spot; they don't understand that ``Clouds are not spheres, mountains are not cones, coastlines are not circles, \ldots"~\cite{Mandelbrot}. They are unaware of the chaos and strange attractors of nonlinear dynamics, where exponentially separating trajectories bounded by finite energy repeatedly stretch and fold into complicated self-similar fractals. Their attempts to learn and predict nonlinear dynamics can be frustrated by ordered and chaotic orbits coexisting at the same energy for different initial positions and momenta.

Recent research~\cite{HNN,HGN} features artificial neural networks that incorporate Hamiltonian structure to learn simple dynamical systems, especially those with outputs proportional to inputs. But from stormy weather to swirling galaxies most dynamical systems are \textit{nonlinear}, exhibit far richer behavior, and pose additional challenges. In this Letter, we exploit the Hamiltonian structure of natural systems to provide neural networks with the physics intelligence needed to learn the mix of order and chaos that often characterizes natural phenomena. After reviewing Hamiltonian chaos and neural networks, we apply Hamiltonian neural networks to the H\'enon-Heiles model, which describes both stellar~\cite{henon,galactic_dynamics_HH} and molecular~\cite{Boyd1981,Feit1984,Vendrell2011} dynamics. Even as these systems transition from order to chaos, Hamiltonian neural networks correctly predict their dynamics, overcoming deep learning's chaos blindness. Physics thereby enhances neural networks, and physics savvy neural networks in turn will help scientists solve hard problems.
 
\begin{figure}[t!]
	\includegraphics[width=0.9\linewidth]{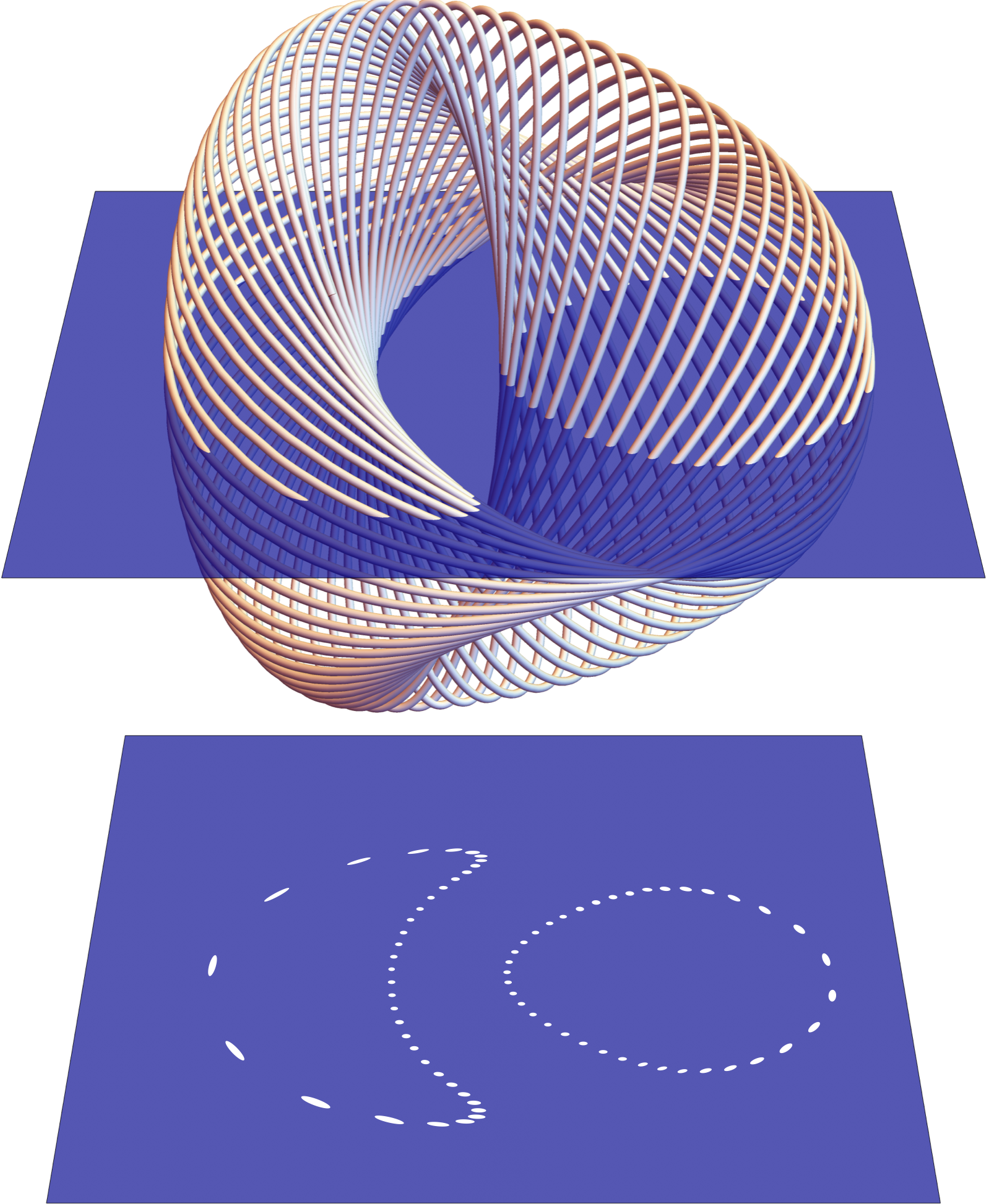} 
	\caption{Flow to section. Phase or state space motion of integrable Hamiltonian systems are confined to invariant tori, like the two-dimensional torus in the four-dimensional phase space of a small-amplitude low-energy H\'enon-Heiles flow (top). Cutting the torus with a plane creates a Poincar\'e surface of section (bottom).}
	\label{FlowToSection}
\end{figure}

% %%%%%%%%%%%%%%%%%%%%%%%%%%%%%%

\textit{Hamiltonian chaos}.---The Hamiltonian formalism describes phenomena  from astronomical scales to nanoscales. Even dissipative systems involving friction or viscosity are microscopically Hamiltonian. It reveals underlying structures in position-momentum phase space and reflects essential symmetries in physical systems. Its elegance stems from its geometric structure, where positions $q$ and conjugate momenta $p$ form a set of $2N$ canonical coordinates describing a physical system with $N$ degrees of freedom. A {\em single} Hamiltonian function $\mathcal{H}[q, p, t]$ uniquely determines the time evolution of the system via the $2N$ coupled differential equations
\begin{equation} \label{hamEq}
      \left\{
     \dot q,
    \dot p
    \right\}
    =
    \left\{
    \mathrm{d} q / \mathrm{d} t,
    \mathrm{d} p / \mathrm{d} t
    \right\}
    =
     \left\{
     +\partial {\mathcal {H}} / \partial { p},
     -\partial {\mathcal {H}} / \partial { q} 
     \right\},
\end{equation}
where the overdots are Newton's notation for time derivatives. For a time-independent system, the total energy $E = \mathcal{H}[q, p]$, which for simple systems is the sum of the kinetic and the potential energies. 

This classical formalism exhibits two contrasting motions. One is simple predictable near-integrable motion that suggests a ``clockwork universe”. Additional motion constants constrain the orbits to lie on low-dimensional Kolmogorov-Arnold-Moser (KAM) tori~\cite{moser1973stable} of dimension $N$ in the 2$N$-dimensional phase space, as in Fig.~\ref{FlowToSection}. Too much nonlinearity can cause adjacent smooth KAM-tori to “break up” into infinitely intersecting fractal cantori~\cite{MACKAY198455}, allowing the orbits to wander over the entire available phase space, constrained only by energy. Such systems can thereby also exhibit chaos where the dynamics, though deterministic, is practically unpredictable due to the extreme sensitivity to initial conditions.

 \begin{figure}[t!]
    \centering
    \includegraphics[width=0.9\linewidth,angle=0]{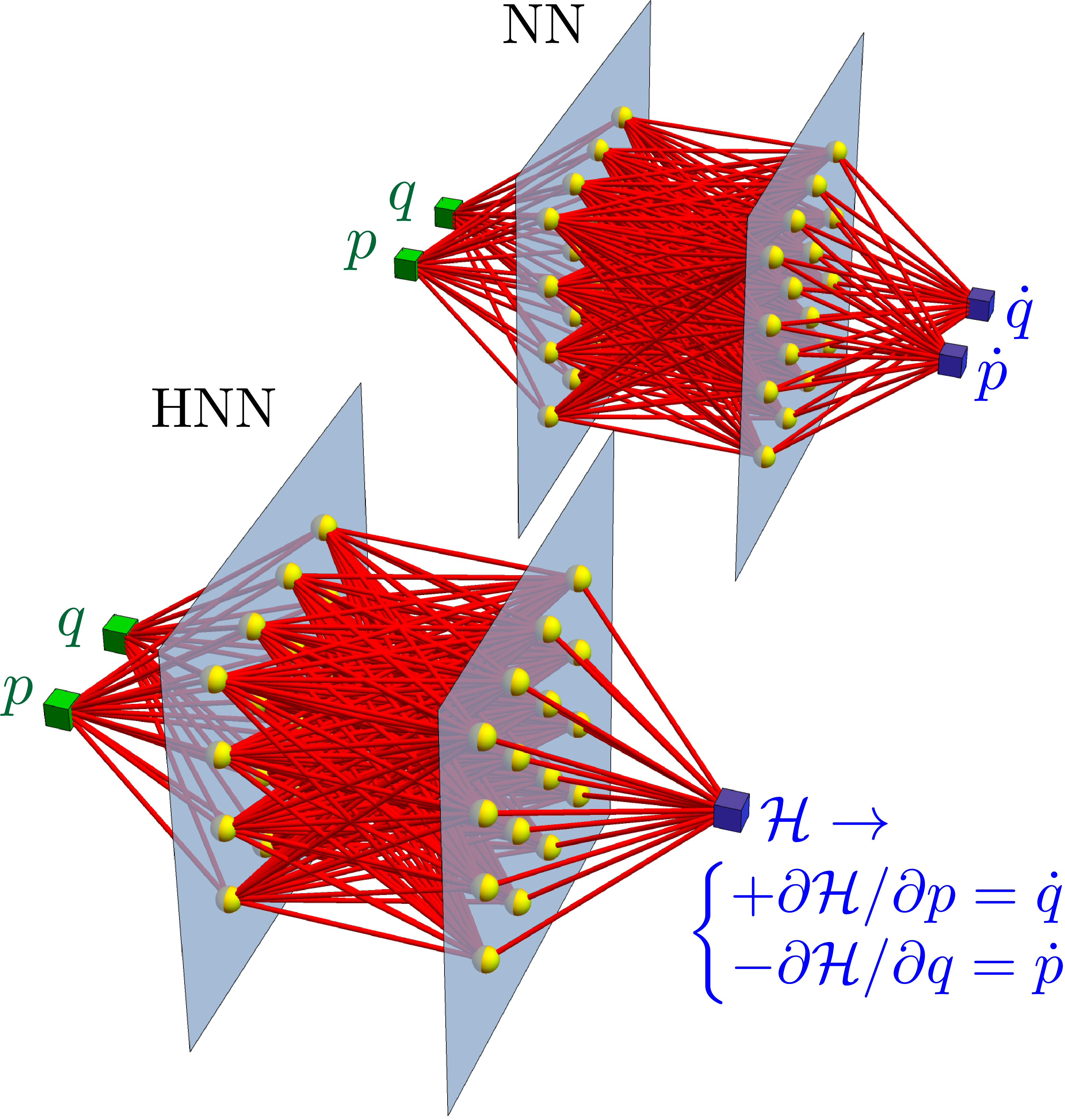}
    \caption{Neural network schematics. Weights (red lines) and biases (yellow spheres) connect inputs (green cubes) through neuron hidden layers (gray planes) to outputs (blue cubes). NN (top) has $2N$ inputs and $2N$ outputs. HNN (bottom) has $2N$ inputs and $1$ output but internalizes the output's gradient in its weights and biases.}
    \label{fig:schematic_HNN}
\end{figure}

% %%%%%%%%%%%%%%%%%%%%%%%%%%%%%%
%\textit{H\'enon-Heiles}.---
From the motion of stars around galactic centers to the vibrations of triatomic molecules, the Hénon-Heiles Hamiltonian~\cite{henon} is a celebrated paradigm of nonlinear dynamics. It exhibits a transition between order and chaos via a mixed phase space where islands of order are embedded in a sea of chaos, one of the most challenging dynamical scenarios to identify and decipher. In a four-dimensional phase space $\{q,p\} = \{q_x,q_y,p_x,p_y\}$, its nondimensionalized Hamiltonian
\begin{equation}
   \mathcal{H}=\left(p_{x}^{2}+p_{y}^{2}\right)/2 + \left(q_x^{2}+q_y^{2}\right)/2+ \left(q_x^2 q_y-{q_y^3/3}\right)
\end{equation}
is the sum of the kinetic and potential energies, including quadratic harmonic terms perturbed by cubic nonlinearities that convert a circularly symmetric potential into a triangularly symmetric potential. Bounded motion is possible in a triangular region of the $\{x,y\}$ plane for energies $0 < E < 1/6$. Figure~\ref{FlowToSection} shows a low energy orbit on a KAM torus and the corresponding Poincar\'e surface of section.

% %%%%%%%%%%%%%%%%%%%%%%%%%%%%%%
%\section{\label{sec:MLP}Baseline Neural Network (NN)}

%\section{Neural Networks}
\textit{Neural networks}.---While traditional analyses focuses on forecasting orbits or understanding fractal structure, understanding the entire landscape of dynamical order and chaos requires new tools. Artificial neural networks are today widely used and studied partly because they can approximate any continuous function~\cite{Cybenko1989, Hornik1991}. Recent efforts to apply them to chaotic dynamics involve the \textit{recurrent} neural networks of \textit{reservoir computing}~\cite{Jaeger78,Pathak2018,Carroll2019}. We instead exploit  the dominant \textit{feed-forward} neural networks of \textit{deep learning}~\cite{Haykin2008}.

Inspired by natural neural networks, the activity $a_\ell =\sigma [W_\ell\,  a_{\ell-1} + b_\ell]$ of each layer of a conventional feed-forward neural network is the nonlinear step or ramp of the linearly transformed activities of the previous layer, where $\sigma$ is a vectorized nonlinear function that mimics the on-off activity of a natural neuron, $a_\ell$ are activation vectors, and $W_\ell$ and $b_\ell$ are adjustable weight matrices and bias vectors that mimic the dendrite and axon connectivity of natural neurons. Concatenating multiple layers eliminates the hidden neuron activities, so the output $y = f[x; W_\ell, b_\ell]$ is a nonlinear function of just the input $x$ and the weights and biases. A training session inputs multiple $x$ and adjusts the weights and biases to minimize the difference or ``loss” $\mathcal{L} =(y_t - y)^2$ between the target $y_t$ and the output $y$ so the neural network learns the correspondence.

\begin{figure*}[ht]
    \centering
	\includegraphics[width=1.0\linewidth]{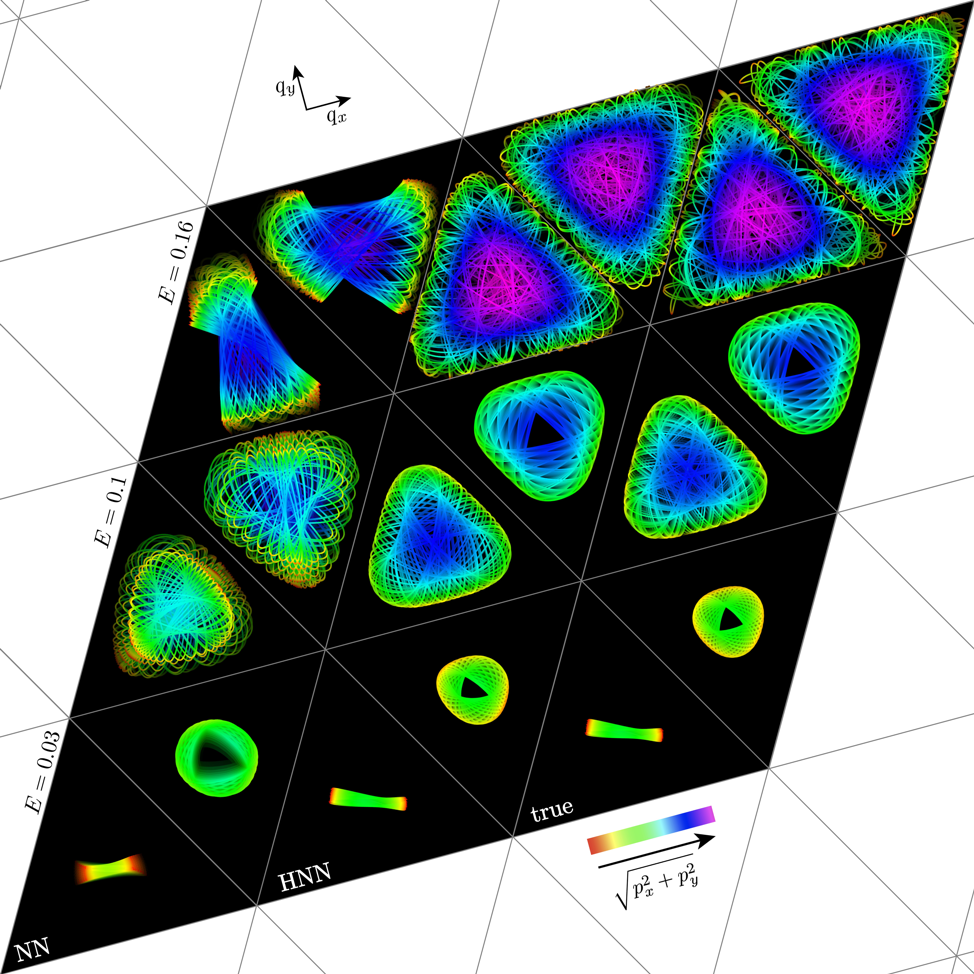}
	\caption{Order-to-chaos flows. Sample H\'enon-Heiles $\{q_x,q_y\}$ flows from conventional neural network (left), Hamiltonian neural network (center), and H\'enon-Heiles differential equations (right), for small, medium, and large bounded energies $0<E<1/6$. Hues code momentum magnitudes, from red to violet; orbit tangents code momentum directions. Orbits fade into the past. HNN has learned which types of orbits are appropriate to which energies. NN is especially poor at high energies.}
	\label{FlowsComparison}
\end{figure*}

% %%%%%%%%%%%%%%%%%%%%%%%%%%%%%%
%\section{\label{sec:HNN} Hamiltonian Neural Network (HNN)}

Recently, neural networks have been proposed~\cite{HNN,HGN} that not only learn the dynamics of the system but also capture invariants and symmetries of the system, including its Hamiltonian phase space structure. The Fig.~\ref{fig:schematic_HNN} conventional neural network NN intakes positions and momenta $\{q, p\}$ and outputs approximations to their time derivatives $\{\dot q, \dot p\}$, adjusting its weights and biases to minimize the loss
\begin{equation} \label{NNLossEq}
	\mathcal{L}_{\text{NN}} = (\dot{q}_t-\dot{q})^2 + (\dot{p}_t-\dot{p})^2
\end{equation}
until it learns the correct mapping. In contrast, the Fig.~\ref{fig:schematic_HNN} Hamiltonian neural network (HNN) intakes position and momenta $\{q, p\}$, outputs the scalar function $\mathcal{H}$, takes its gradient to find its position and momentum rates of change, and minimizes the loss 
\begin{equation}\label{HNNLossEq}
    \mathcal{L}_{\text{HNN}}
    = \left(\dot q_t - \partial\mathcal{H} / \partial p \right)^2
    + \left(\dot p_t + \partial\mathcal{H} / \partial q \right)^2,
\end{equation}
% \begin{equation}\label{HNNLossEq}
%     \mathcal{L}_{\text{HNN}}
%     = \left(\dot q_t - \frac{\partial\mathcal{H}}{\partial p} \right)^2
%     + \left(\dot p_t + \frac{\partial\mathcal{H}}{\partial q} \right)^2,
% \end{equation}
%
which enforces Hamilton's equations of motion. For a given time step $dt$, each trained network can extrapolate a given initial condition with an Euler update $ \{q, p\} \leftarrow \{q, p\} + \{\dot q, \dot p\} dt $
% %
% \begin{equation}
%   \{q, p\} &\leftarrow \{q, p\} + \{\dot q, \dot p\} dt 
% \end{equation}
% %
or some better integration scheme~\cite{NR}.

HNN incorporates the physics bias that the output phase-space velocities $\{\dot q, \dot p\}$ must come from the gradient of an unknown but conserved quantity, the Hamiltonian. This enforces an additional constraint on network weights and biases. Instead of taking the difference of the true output and the network generated output, HNN constructs the Eq.~\ref{hamEq} conservative vector field from the given coordinates by differentiating the output with respect to the inputs. It then uses this gradient to construct the loss function that in turn optimizes the vector field.

\begin{figure}[b!]
    \centering
    \includegraphics[width=1.0\linewidth]{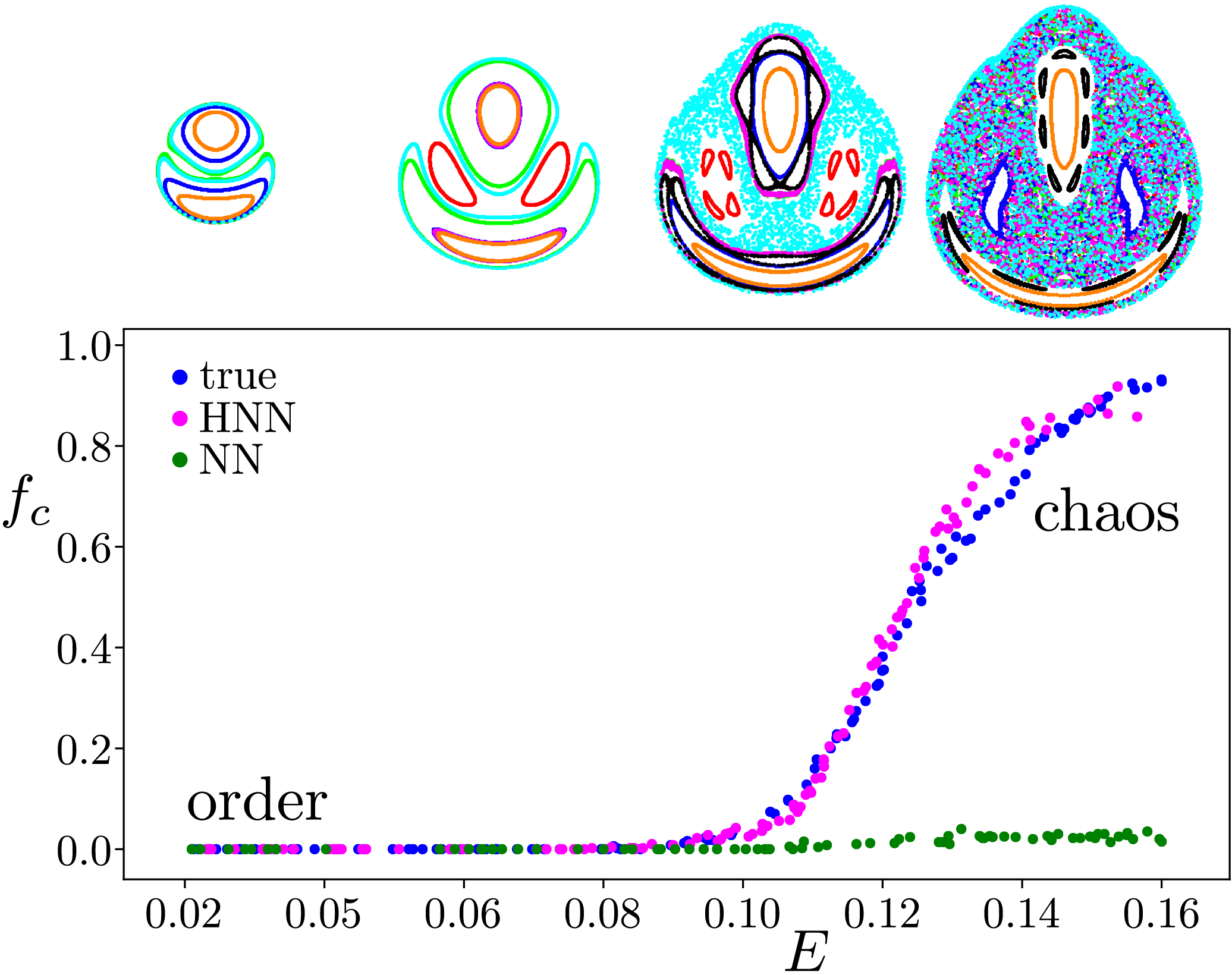}
    \caption{Order-to-chaos sections. Fraction of chaotic orbits $f_c$ for random energies $E$, as inferred by the smaller alignment index $\alpha$, for conventional neural network (green), Hamiltonian neural network (magenta), and H\'enon-Heiles differential equations (blue). True Poincar\'e $\{q_y,p_y\}$ sections (top) illustrate the order-chaos transition, where each color corresponds to the same initial condition at different energies. NN is again especially poor for the mainly chaotic high-energy orbits.}
    \label{fig:sali}
\end{figure}

% %%%%%%%%%%%%%%%%%%%%%%%%%%%%%%
%\section{Results}
\textit{Results}.---For a sample of bounded energies and with the same learning parameters~\cite{Supplement}, we train NN and HNN on multiple H\'enon-Heiles trajectories starting in the triangular basin. We use the neural networks to forecast new trajectories, and then compare them to the ``true" trajectories obtained by numerically integrating Hamilton's Eq.~\ref{hamEq}. Figure~\ref{FlowsComparison} shows these results. HNN captures the nature of the global phase space structures well and effectively distinguishes qualitatively different dynamical regimes. NN fails dramatically, especially at high energies.

To quantify the ability of NN and HNN to paint a full portrait of the global, mixed phase space dynamics, we use their knowledge of the system to estimate the H\'enon-Heiles Lyapunov spectrum~\cite{Strogatz}, which characterizes the separation rate of infinitesimally close trajectories, one exponent for each dimension. Since perturbations along the flow do not cause divergence away from it, at least one exponent will be zero. For a Hamiltonian system, the exponents must exist in diverging-converging pairs to conserve phase space volume. Hence we expect a spectrum like $\{-\lambda,0,0,+\lambda \}$, with the maximum exponent increasing at large energies like $\lambda \propto E^{3.5}$~\cite{Melnikov2004}. HNN satisfies both these expectations, which are stringent non-trivial consistency checks that it has authentically learned the true flow. NN satisfies neither~\cite{Supplement}. 

Using NN and HNN we also compute the smaller alignment index $\alpha$, a metric of chaos that allows us to quickly find the fraction of orbits that are chaotic at any energy~\cite{Skokos_2001}. We compute $\alpha$ for a specific orbit by following the time evolution of two different normalized deviation vectors along the orbit and computing the minimum of the norms of their difference and sum. Via extensive testing, an orbit is chaotic if $\alpha < 10^{-8}$, indicating that its deviation vectors have been aligned or anti-aligned by a large positive Lyapunov exponent. Figure~\ref{fig:sali} shows the fraction of chaotic trajectories for each energy, including a dramatic transition between islands of order at low energy and a sea of chaos at high energy. The chaos fractions computed by numerically integrating Hamilton's Eq.~\ref{hamEq} are similar to the HNN estimates. NN again fails dramatically.

\begin{figure}[ht!]
    \centering
	\includegraphics[width=1.0\linewidth]{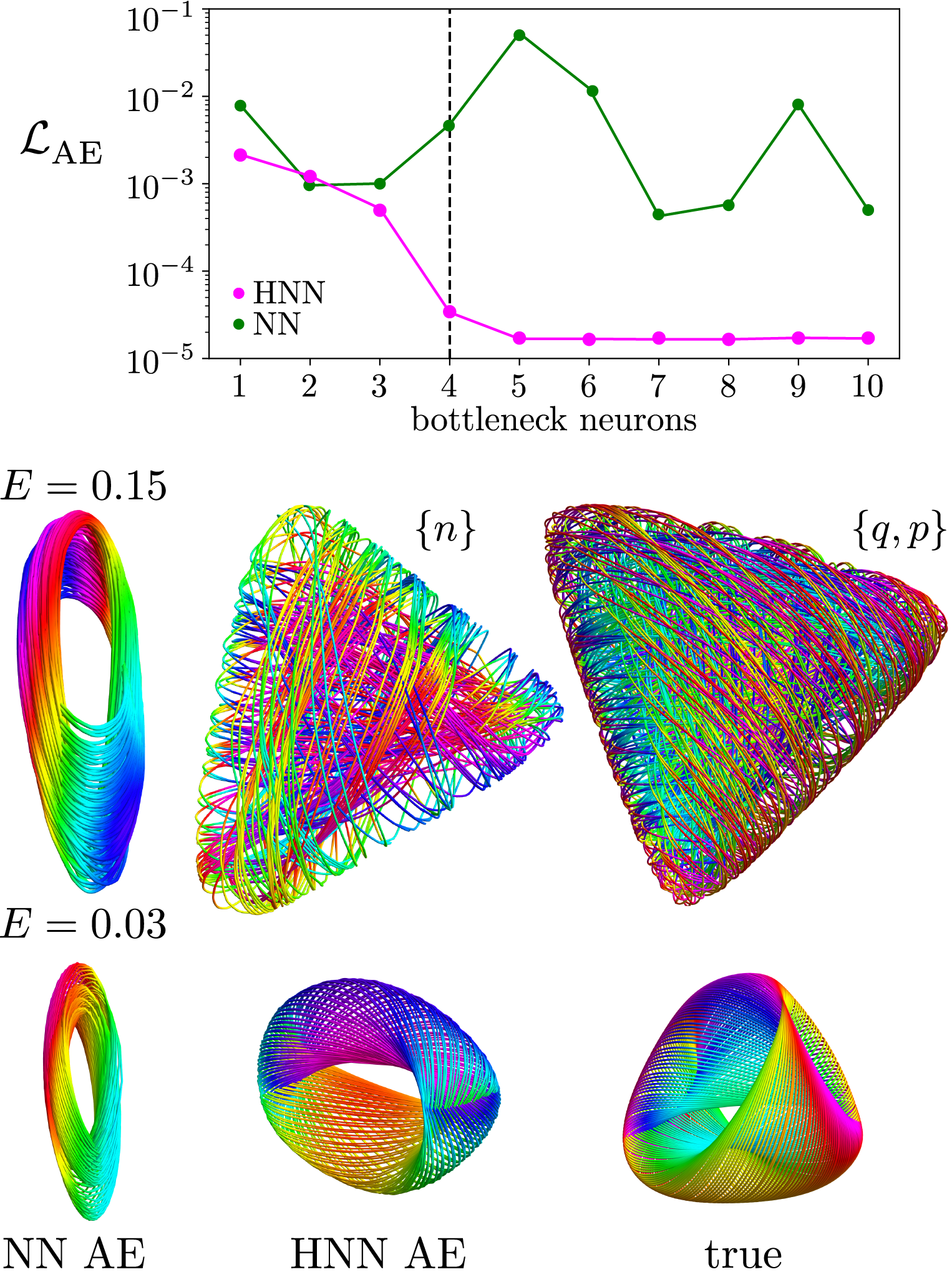}
	\caption{Introspection. When an autoencoder compresses HNN's forecasting activity, its loss function $\mathcal{L_\text{AE}}$ (magenta top) drops precipitously when its bottleneck layer increases past 4 neurons,the dimensionality of the H\'enon-Heiles system. NN seems oblivious (green top) to this transition. HNN's compressed representation $\{n_1,n_2,n_3,n_4\}$ resembles the low or high energy orbit $\{q_x,q_y,p_x,p_y\}$ it is forecasting, where color indicates the 4th dimension (bottom). NN hardly notices the qualitative differences in these orbits. }
	\label{introspection}
\end{figure}

Finally, to understand what NN and HNN have learned when they forecast orbits, we use an \textit{autoencoder} -- a neural network with a sparse ``bottleneck" layer -- to examine their hidden neurons~\cite{Supplement}. The autoencoder's Mean Square Error (MSE) loss function forces the input to match the output, so it must adjust its weights and biases to create a compressed, low-dimensional representation of the neural networks' activity, a process called introspection~\cite{wang2019emergent} or intelligible artificial intelligence (\textit{i}AI). For HNN, the loss function drops precipitously for 4 (or more) bottleneck neurons, which appear to encode a linear combination of the $4$ phase space coordinates, thereby capturing the dimensionality of the system, as in Fig.~\ref{introspection}. NN shows no similar drop, and the uncertainty in its loss function is orders of magnitude larger than HNN's.

% \begin{figure}[ht!]
%     \centering
% 	\includegraphics[width=1.0\linewidth]{figures/InstrospectionLog.png}
% 	\caption{Introspection. When an autoencoder (top) compresses HNN's forecasting activity, its loss function $\mathcal{L_\text{AE}}$ (magenta middle) drops precipitously when its bottleneck layer increases to 4 neurons,the dimensionality of the H\'enon-Heiles system. NN seems oblivious (green middle) to this transition. HNN's compressed representation $\{n_1,n_2,n_3,n_4\}$ resembles the low or high energy orbit $\{q_x,q_y,p_x,p_y\}$ it is forecasting, where color indicates the 4th dimension (bottom). NN hardly notices the qualitative differences in these orbits. }
% 	\label{introspection}
% \end{figure}

% %%%%%%%%%%%%%%%%%%%%%%%%%%%%%%

%\section{Conclusion}
\textit{Conclusion}.---Time series can be analyzed using multiple techniques, including genetic algorithms~\cite{Schmidt81} or neural networks~\cite{AIFeynman} to find algebraic equations of motion or simple compressed representations~\cite{SciNet}. But such systems merely discover underlying equations or relations. Newton and Poincar\'e had the equations hundreds of years ago and still just glimpsed their complexity without fully understanding it. Conventional neural networks extrapolating time series do not conserve energy, and their orbits can drift off the energy surface, jump into the sea of chaos from islands of order, or fly out to infinity. By incorporating energy conserving and volume preserving flows arising from an underlying Hamiltonian function --  \textit{without invoking any details of its form} -- Hamiltonian neural networks can recognize the presence of order and chaos as well as the challenging regime where both these very distinct dynamics coexist. 

A neural network that respects Hamiltonian time-translational symmetry can learn order and chaos, including mixed phase space flows and sections, as quantified by metrics like Lyapunov spectra and smaller alignment indices. Incorporating other symmetries~\cite{Bondesan2019} in deep learning may produce comparable qualitative performance improvements. We are excited about the potential for physics to improve artificial neural networks, which can return the favor by helping us solve hard problems. Future researchers will work alongside such artificial intelligences to help extend the frontiers of science.

This research was supported by ONR grant N00014-16-1-3066 and a gift from United Therapeutics. J.F.L. thanks The College of Wooster for making possible his sabbatical at NCSU. S.S. acknowledges support from the J.C. Bose National Fellowship (SB/S2/JCB-013/2015).

% %%%%%%%%%%%%%%%%%%%%%%%%%%%%%%
\bibliography{LearningChaos}% Produces the bibliography via BibTeX.

\providecommand{\noopsort}[1]{}\providecommand{\singleletter}[1]{#1}%
\begin{thebibliography}{36}%
\makeatletter
\providecommand \@ifxundefined [1]{%
 \@ifx{#1\undefined}
}%
\providecommand \@ifnum [1]{%
 \ifnum #1\expandafter \@firstoftwo
 \else \expandafter \@secondoftwo
 \fi
}%
\providecommand \@ifx [1]{%
 \ifx #1\expandafter \@firstoftwo
 \else \expandafter \@secondoftwo
 \fi
}%
\providecommand \natexlab [1]{#1}%
\providecommand \enquote  [1]{``#1''}%
\providecommand \bibnamefont  [1]{#1}%
\providecommand \bibfnamefont [1]{#1}%
\providecommand \citenamefont [1]{#1}%
\providecommand \href@noop [0]{\@secondoftwo}%
\providecommand \href [0]{\begingroup \@sanitize@url \@href}%
\providecommand \@href[1]{\@@startlink{#1}\@@href}%
\providecommand \@@href[1]{\endgroup#1\@@endlink}%
\providecommand \@sanitize@url [0]{\catcode `\\12\catcode `\$12\catcode
  `\&12\catcode `\#12\catcode `\^12\catcode `\_12\catcode `\%12\relax}%
\providecommand \@@startlink[1]{}%
\providecommand \@@endlink[0]{}%
\providecommand \url  [0]{\begingroup\@sanitize@url \@url }%
\providecommand \@url [1]{\endgroup\@href {#1}{\urlprefix }}%
\providecommand \urlprefix  [0]{URL }%
\providecommand \Eprint [0]{\href }%
\providecommand \doibase [0]{https://doi.org/}%
\providecommand \selectlanguage [0]{\@gobble}%
\providecommand \bibinfo  [0]{\@secondoftwo}%
\providecommand \bibfield  [0]{\@secondoftwo}%
\providecommand \translation [1]{[#1]}%
\providecommand \BibitemOpen [0]{}%
\providecommand \bibitemStop [0]{}%
\providecommand \bibitemNoStop [0]{.\EOS\space}%
\providecommand \EOS [0]{\spacefactor3000\relax}%
\providecommand \BibitemShut  [1]{\csname bibitem#1\endcsname}%
\let\auto@bib@innerbib\@empty
%</preamble>
\bibitem [{\citenamefont {Szebehely}\ and\ \citenamefont
  {Mark}(1998)}]{NewtonQuote}%
  \BibitemOpen
  \bibfield  {author} {\bibinfo {author} {\bibfnamefont {V.~G.}\ \bibnamefont
  {Szebehely}}\ and\ \bibinfo {author} {\bibfnamefont {H.}~\bibnamefont
  {Mark}},\ }\href@noop {} {\emph {\bibinfo {title} {Adventures in Celestial
  Mechanics}}}\ (\bibinfo  {publisher} {Second Edition. John Wiley \& Sons},\
  \bibinfo {year} {1998})\BibitemShut {NoStop}%
\bibitem [{\citenamefont {Hamilton}(1834)}]{Hamilton}%
  \BibitemOpen
  \bibfield  {author} {\bibinfo {author} {\bibfnamefont {W.~R.}\ \bibnamefont
  {Hamilton}},\ }\bibfield  {title} {\bibinfo {title} {{O}n a general method in
  dynamics; by which the study of the motions of all free systems of attracting
  or repelling points is reduced to the search and differentiation of one
  central relation, or characteristic function},\ }\href@noop {} {\bibfield
  {journal} {\bibinfo  {journal} {Physical Transactions of the Royal Society}\
  }\textbf {\bibinfo {volume} {124}},\ \bibinfo {pages} {247} (\bibinfo {year}
  {1834})}\BibitemShut {NoStop}%
\bibitem [{\citenamefont {Poincar{\'e}}(1899)}]{Poincare}%
  \BibitemOpen
  \bibfield  {author} {\bibinfo {author} {\bibfnamefont {H.}~\bibnamefont
  {Poincar{\'e}}},\ }\bibfield  {title} {\bibinfo {title} {Les m{\'e}thodes
  nouvelles de la m{\'e}canique c{\'e}leste},\ }\href
  {https://doi.org/10.1007/BF02742713} {\bibfield  {journal} {\bibinfo
  {journal} {Il Nuovo Cimento}\ }\textbf {\bibinfo {volume} {10}},\ \bibinfo
  {pages} {128} (\bibinfo {year} {1899})}\BibitemShut {NoStop}%
\bibitem [{\citenamefont {Lorenz}(1963)}]{Lorenz}%
  \BibitemOpen
  \bibfield  {author} {\bibinfo {author} {\bibfnamefont {E.~N.}\ \bibnamefont
  {Lorenz}},\ }\bibfield  {title} {\bibinfo {title} {Deterministic nonperiodic
  flow},\ }\href {https://doi.org/10.1175/1520-0469(1963)020<0130:DNF>2.0.CO;2}
  {\bibfield  {journal} {\bibinfo  {journal} {Journal of the Atmospheric
  Sciences}\ }\textbf {\bibinfo {volume} {20}},\ \bibinfo {pages} {130}
  (\bibinfo {year} {1963})}\BibitemShut {NoStop}%
\bibitem [{\citenamefont {Gleick}(1987)}]{Gleick}%
  \BibitemOpen
  \bibfield  {author} {\bibinfo {author} {\bibfnamefont {J.}~\bibnamefont
  {Gleick}},\ }\href@noop {} {\emph {\bibinfo {title} {Chaos: Making a New
  Science}}}\ (\bibinfo  {publisher} {Penguin Books},\ \bibinfo {address} {New
  York},\ \bibinfo {year} {1987})\BibitemShut {NoStop}%
\bibitem [{\citenamefont {Haykin}(0008)}]{Haykin2008}%
  \BibitemOpen
  \bibfield  {author} {\bibinfo {author} {\bibfnamefont {S.~O.}\ \bibnamefont
  {Haykin}},\ }\href@noop {} {\emph {\bibinfo {title} {Neural Networks and
  Learning Machines}}}\ (\bibinfo  {publisher} {Third edition. Pearson},\
  \bibinfo {year} {20008})\BibitemShut {NoStop}%
\bibitem [{\citenamefont {Lusch}\ \emph {et~al.}(2018)\citenamefont {Lusch},
  \citenamefont {Kutz},\ and\ \citenamefont {Brunton}}]{Lusch2018}%
  \BibitemOpen
  \bibfield  {author} {\bibinfo {author} {\bibfnamefont {B.}~\bibnamefont
  {Lusch}}, \bibinfo {author} {\bibfnamefont {J.~N.}\ \bibnamefont {Kutz}},\
  and\ \bibinfo {author} {\bibfnamefont {S.~L.}\ \bibnamefont {Brunton}},\
  }\bibfield  {title} {\bibinfo {title} {Deep learning for universal linear
  embeddings of nonlinear dynamics},\ }\href
  {https://doi.org/10.1038/s41467-018-07210-0} {\bibfield  {journal} {\bibinfo
  {journal} {Nature Communications}\ }\textbf {\bibinfo {volume} {9}},\
  \bibinfo {pages} {4950} (\bibinfo {year} {2018})}\BibitemShut {NoStop}%
\bibitem [{\citenamefont {Iten}\ \emph {et~al.}(2019)\citenamefont {Iten},
  \citenamefont {Metger}, \citenamefont {Wilming}, \citenamefont {del Rio},\
  and\ \citenamefont {Renner}}]{SciNet}%
  \BibitemOpen
  \bibfield  {author} {\bibinfo {author} {\bibfnamefont {R.}~\bibnamefont
  {Iten}}, \bibinfo {author} {\bibfnamefont {T.}~\bibnamefont {Metger}},
  \bibinfo {author} {\bibfnamefont {H.}~\bibnamefont {Wilming}}, \bibinfo
  {author} {\bibfnamefont {L.}~\bibnamefont {del Rio}},\ and\ \bibinfo {author}
  {\bibfnamefont {R.}~\bibnamefont {Renner}},\ }\bibfield  {title} {\bibinfo
  {title} {Discovering physical concepts with neural networks},\ }\href@noop {}
  {\bibfield  {journal} {\bibinfo  {journal} {Phys. Rev. Lett.}\ }\textbf
  {\bibinfo {volume} {TBA}},\ \bibinfo {pages} {TBA} (\bibinfo {year}
  {2019})}\BibitemShut {NoStop}%
\bibitem [{\citenamefont {Udrescu}\ and\ \citenamefont
  {Tegmark}(2019)}]{AIFeynman}%
  \BibitemOpen
  \bibfield  {author} {\bibinfo {author} {\bibfnamefont {S.-M.}\ \bibnamefont
  {Udrescu}}\ and\ \bibinfo {author} {\bibfnamefont {M.}~\bibnamefont
  {Tegmark}},\ }\bibfield  {title} {\bibinfo {title} {{AI} {F}eynman: a
  {P}hysics-{I}nspired {M}ethod for {S}ymbolic {R}egression},\ }\href@noop {}
  {\bibfield  {journal} {\bibinfo  {journal} {arXiv:1905.11481}\ } (\bibinfo
  {year} {2019})}\BibitemShut {NoStop}%
\bibitem [{\citenamefont {Wu}\ and\ \citenamefont
  {Tegmark}(2019)}]{AIPhysicist}%
  \BibitemOpen
  \bibfield  {author} {\bibinfo {author} {\bibfnamefont {T.}~\bibnamefont
  {Wu}}\ and\ \bibinfo {author} {\bibfnamefont {M.}~\bibnamefont {Tegmark}},\
  }\bibfield  {title} {\bibinfo {title} {Toward an artificial intelligence
  physicist for unsupervised learning},\ }\href
  {https://doi.org/10.1103/PhysRevE.100.033311} {\bibfield  {journal} {\bibinfo
   {journal} {Phys. Rev. E}\ }\textbf {\bibinfo {volume} {100}},\ \bibinfo
  {pages} {033311} (\bibinfo {year} {2019})}\BibitemShut {NoStop}%
\bibitem [{\citenamefont {Anonymous}(2020)}]{anonymous2020deep}%
  \BibitemOpen
  \bibfield  {author} {\bibinfo {author} {\bibnamefont {Anonymous}},\
  }\bibfield  {title} {\bibinfo {title} {Deep learning for symbolic
  mathematics},\ }in\ \href {https://openreview.net/forum?id=S1eZYeHFDS} {\emph
  {\bibinfo {booktitle} {Submitted to International Conference on Learning
  Representations}}}\ (\bibinfo {year} {2020})\ \bibinfo {note} {under
  review}\BibitemShut {NoStop}%
\bibitem [{\citenamefont {Breen}\ \emph {et~al.}(2019)\citenamefont {Breen},
  \citenamefont {Foley}, \citenamefont {Boekholt},\ and\ \citenamefont
  {Zwart}}]{breen2019newton}%
  \BibitemOpen
  \bibfield  {author} {\bibinfo {author} {\bibfnamefont {P.~G.}\ \bibnamefont
  {Breen}}, \bibinfo {author} {\bibfnamefont {C.~N.}\ \bibnamefont {Foley}},
  \bibinfo {author} {\bibfnamefont {T.}~\bibnamefont {Boekholt}},\ and\
  \bibinfo {author} {\bibfnamefont {S.~P.}\ \bibnamefont {Zwart}},\ }\bibfield
  {title} {\bibinfo {title} {Newton vs the machine: solving the chaotic
  three-body problem using deep neural networks},\ }\href@noop {} {\bibfield
  {journal} {\bibinfo  {journal} {arXiv:1910.07291}\ } (\bibinfo {year}
  {2019})}\BibitemShut {NoStop}%
\bibitem [{\citenamefont {Silver}\ \emph {et~al.}(2018)\citenamefont {Silver},
  \citenamefont {Hubert}, \citenamefont {Schrittwieser}, \citenamefont
  {Antonoglou}, \citenamefont {Lai}, \citenamefont {Guez}, \citenamefont
  {Lanctot}, \citenamefont {Sifre}, \citenamefont {Kumaran}, \citenamefont
  {Graepel}, \citenamefont {Lillicrap}, \citenamefont {Simonyan},\ and\
  \citenamefont {Hassabis}}]{AlphaZero}%
  \BibitemOpen
  \bibfield  {author} {\bibinfo {author} {\bibfnamefont {D.}~\bibnamefont
  {Silver}}, \bibinfo {author} {\bibfnamefont {T.}~\bibnamefont {Hubert}},
  \bibinfo {author} {\bibfnamefont {J.}~\bibnamefont {Schrittwieser}}, \bibinfo
  {author} {\bibfnamefont {I.}~\bibnamefont {Antonoglou}}, \bibinfo {author}
  {\bibfnamefont {M.}~\bibnamefont {Lai}}, \bibinfo {author} {\bibfnamefont
  {A.}~\bibnamefont {Guez}}, \bibinfo {author} {\bibfnamefont {M.}~\bibnamefont
  {Lanctot}}, \bibinfo {author} {\bibfnamefont {L.}~\bibnamefont {Sifre}},
  \bibinfo {author} {\bibfnamefont {D.}~\bibnamefont {Kumaran}}, \bibinfo
  {author} {\bibfnamefont {T.}~\bibnamefont {Graepel}}, \bibinfo {author}
  {\bibfnamefont {T.}~\bibnamefont {Lillicrap}}, \bibinfo {author}
  {\bibfnamefont {K.}~\bibnamefont {Simonyan}},\ and\ \bibinfo {author}
  {\bibfnamefont {D.}~\bibnamefont {Hassabis}},\ }\bibfield  {title} {\bibinfo
  {title} {A general reinforcement learning algorithm that masters chess,
  shogi, and go through self-play},\ }\href
  {https://doi.org/10.1126/science.aar6404} {\bibfield  {journal} {\bibinfo
  {journal} {Science}\ }\textbf {\bibinfo {volume} {362}},\ \bibinfo {pages}
  {1140} (\bibinfo {year} {2018})}\BibitemShut {NoStop}%
\bibitem [{\citenamefont {Mandelbrot}(1982)}]{Mandelbrot}%
  \BibitemOpen
  \bibfield  {author} {\bibinfo {author} {\bibfnamefont {B.~B.}\ \bibnamefont
  {Mandelbrot}},\ }\href {https://cds.cern.ch/record/98509} {\emph {\bibinfo
  {title} {{The fractal geometry of nature}}}}\ (\bibinfo  {publisher}
  {Freeman},\ \bibinfo {address} {San Francisco, CA},\ \bibinfo {year}
  {1982})\BibitemShut {NoStop}%
\bibitem [{\citenamefont {Greydanus}\ \emph {et~al.}(2019)\citenamefont
  {Greydanus}, \citenamefont {Dzamba},\ and\ \citenamefont {Yosinski}}]{HNN}%
  \BibitemOpen
  \bibfield  {author} {\bibinfo {author} {\bibfnamefont {S.}~\bibnamefont
  {Greydanus}}, \bibinfo {author} {\bibfnamefont {M.}~\bibnamefont {Dzamba}},\
  and\ \bibinfo {author} {\bibfnamefont {J.}~\bibnamefont {Yosinski}},\
  }\bibfield  {title} {\bibinfo {title} {Hamiltonian neural networks},\
  }\href@noop {} {\bibfield  {journal} {\bibinfo  {journal} {arXiv:1906.01563}\
  } (\bibinfo {year} {2019})}\BibitemShut {NoStop}%
\bibitem [{\citenamefont {Toth}\ \emph {et~al.}(2019)\citenamefont {Toth},
  \citenamefont {Rezende}, \citenamefont {Jaegle}, \citenamefont {Racanière},
  \citenamefont {Botev},\ and\ \citenamefont {Higgins}}]{HGN}%
  \BibitemOpen
  \bibfield  {author} {\bibinfo {author} {\bibfnamefont {P.}~\bibnamefont
  {Toth}}, \bibinfo {author} {\bibfnamefont {D.~J.}\ \bibnamefont {Rezende}},
  \bibinfo {author} {\bibfnamefont {A.}~\bibnamefont {Jaegle}}, \bibinfo
  {author} {\bibfnamefont {S.}~\bibnamefont {Racanière}}, \bibinfo {author}
  {\bibfnamefont {A.}~\bibnamefont {Botev}},\ and\ \bibinfo {author}
  {\bibfnamefont {I.}~\bibnamefont {Higgins}},\ }\bibfield  {title} {\bibinfo
  {title} {Hamiltonian generative networks},\ }\href@noop {} {\bibfield
  {journal} {\bibinfo  {journal} {arXiv:1909.13789}\ } (\bibinfo {year}
  {2019})}\BibitemShut {NoStop}%
\bibitem [{\citenamefont {{H\'enon}}\ and\ \citenamefont
  {{Heiles}}(1964)}]{henon}%
  \BibitemOpen
  \bibfield  {author} {\bibinfo {author} {\bibfnamefont {M.}~\bibnamefont
  {{H\'enon}}}\ and\ \bibinfo {author} {\bibfnamefont {C.}~\bibnamefont
  {{Heiles}}},\ }\bibfield  {title} {\bibinfo {title} {{The applicability of
  the third integral of motion: Some numerical experiments}},\ }\href
  {https://doi.org/10.1086/109234} {\bibfield  {journal} {\bibinfo  {journal}
  {Astronomical Journal}\ }\textbf {\bibinfo {volume} {69}},\ \bibinfo {pages}
  {73} (\bibinfo {year} {1964})}\BibitemShut {NoStop}%
\bibitem [{\citenamefont {Binney}\ and\ \citenamefont
  {Tremaine}(2011)}]{galactic_dynamics_HH}%
  \BibitemOpen
  \bibfield  {author} {\bibinfo {author} {\bibfnamefont {J.}~\bibnamefont
  {Binney}}\ and\ \bibinfo {author} {\bibfnamefont {S.}~\bibnamefont
  {Tremaine}},\ }\href@noop {} {\emph {\bibinfo {title} {Galactic
  {d}ynamics}}},\ Vol.~\bibinfo {volume} {20}\ (\bibinfo  {publisher}
  {Princeton university press},\ \bibinfo {year} {2011})\BibitemShut {NoStop}%
\bibitem [{\citenamefont {Waite}\ and\ \citenamefont
  {Miller}(1981)}]{Boyd1981}%
  \BibitemOpen
  \bibfield  {author} {\bibinfo {author} {\bibfnamefont {B.~A.}\ \bibnamefont
  {Waite}}\ and\ \bibinfo {author} {\bibfnamefont {W.~H.}\ \bibnamefont
  {Miller}},\ }\bibfield  {title} {\bibinfo {title} {Mode specificity in
  unimolecular reaction dynamics: The {H}\'enon-{H}eiles potential energy
  surface},\ }\href@noop {} {\bibfield  {journal} {\bibinfo  {journal} {The
  Journal of Chemical Physics}\ }\textbf {\bibinfo {volume} {74}},\ \bibinfo
  {pages} {3910} (\bibinfo {year} {1981})}\BibitemShut {NoStop}%
\bibitem [{\citenamefont {Feit}\ and\ \citenamefont {Fleck}(1984)}]{Feit1984}%
  \BibitemOpen
  \bibfield  {author} {\bibinfo {author} {\bibfnamefont {M.~D.}\ \bibnamefont
  {Feit}}\ and\ \bibinfo {author} {\bibfnamefont {J.~A.}\ \bibnamefont
  {Fleck}},\ }\bibfield  {title} {\bibinfo {title} {Wave packet dynamics and
  chaos in the {H}\'enon-{H}eiles system},\ }\href@noop {} {\bibfield
  {journal} {\bibinfo  {journal} {The Journal of Chemical Physics}\ }\textbf
  {\bibinfo {volume} {80}},\ \bibinfo {pages} {2578} (\bibinfo {year}
  {1984})}\BibitemShut {NoStop}%
\bibitem [{\citenamefont {Vendrell}\ and\ \citenamefont
  {Meyer}(2011)}]{Vendrell2011}%
  \BibitemOpen
  \bibfield  {author} {\bibinfo {author} {\bibfnamefont {O.}~\bibnamefont
  {Vendrell}}\ and\ \bibinfo {author} {\bibfnamefont {H.-D.}\ \bibnamefont
  {Meyer}},\ }\bibfield  {title} {\bibinfo {title} {Multilayer
  multiconfiguration time-dependent hartree method: Implementation and
  applications to a {H}\'enon-{H}eiles hamiltonian and to pyrazine},\
  }\href@noop {} {\bibfield  {journal} {\bibinfo  {journal} {The Journal of
  Chemical Physics}\ }\textbf {\bibinfo {volume} {134}},\ \bibinfo {pages}
  {044135} (\bibinfo {year} {2011})}\BibitemShut {NoStop}%
\bibitem [{\citenamefont {Moser}(1973)}]{moser1973stable}%
  \BibitemOpen
  \bibfield  {author} {\bibinfo {author} {\bibfnamefont {J.}~\bibnamefont
  {Moser}},\ }\href {https://books.google.com/books?id=hZCgKQEACAAJ} {\emph
  {\bibinfo {title} {Stable and Random Motions in Dynamical Systems: with
  Special Emphasis on Celestial Mechanics}}},\ Annals of Mathematics studies\
  (\bibinfo  {publisher} {Princeton University Press},\ \bibinfo {year}
  {1973})\BibitemShut {NoStop}%
\bibitem [{\citenamefont {Mackay}\ \emph {et~al.}(1984)\citenamefont {Mackay},
  \citenamefont {Meiss},\ and\ \citenamefont {Percival}}]{MACKAY198455}%
  \BibitemOpen
  \bibfield  {author} {\bibinfo {author} {\bibfnamefont {R.}~\bibnamefont
  {Mackay}}, \bibinfo {author} {\bibfnamefont {J.}~\bibnamefont {Meiss}},\ and\
  \bibinfo {author} {\bibfnamefont {I.}~\bibnamefont {Percival}},\ }\bibfield
  {title} {\bibinfo {title} {Transport in hamiltonian systems},\ }\href
  {https://doi.org/https://doi.org/10.1016/0167-2789(84)90270-7} {\bibfield
  {journal} {\bibinfo  {journal} {Physica D: Nonlinear Phenomena}\ }\textbf
  {\bibinfo {volume} {13}},\ \bibinfo {pages} {55 } (\bibinfo {year}
  {1984})}\BibitemShut {NoStop}%
\bibitem [{\citenamefont {Cybenko}(1989)}]{Cybenko1989}%
  \BibitemOpen
  \bibfield  {author} {\bibinfo {author} {\bibfnamefont {G.}~\bibnamefont
  {Cybenko}},\ }\bibfield  {title} {\bibinfo {title} {{Approximation by
  superpositions of a sigmoidal function}},\ }\href
  {https://doi.org/10.1007/BF02551274} {\bibfield  {journal} {\bibinfo
  {journal} {Mathematics of Control, Signals, and Systems (MCSS)}\ }\textbf
  {\bibinfo {volume} {2}},\ \bibinfo {pages} {303} (\bibinfo {year}
  {1989})}\BibitemShut {NoStop}%
\bibitem [{\citenamefont {Hornik}(1991)}]{Hornik1991}%
  \BibitemOpen
  \bibfield  {author} {\bibinfo {author} {\bibfnamefont {K.}~\bibnamefont
  {Hornik}},\ }\bibfield  {title} {\bibinfo {title} {Approximation capabilities
  of multilayer feedforward networks},\ }\href
  {https://doi.org/https://doi.org/10.1016/0893-6080(91)90009-T} {\bibfield
  {journal} {\bibinfo  {journal} {Neural Networks}\ }\textbf {\bibinfo {volume}
  {4}},\ \bibinfo {pages} {251} (\bibinfo {year} {1991})}\BibitemShut {NoStop}%
\bibitem [{\citenamefont {Jaeger}\ and\ \citenamefont {Haas}(2004)}]{Jaeger78}%
  \BibitemOpen
  \bibfield  {author} {\bibinfo {author} {\bibfnamefont {H.}~\bibnamefont
  {Jaeger}}\ and\ \bibinfo {author} {\bibfnamefont {H.}~\bibnamefont {Haas}},\
  }\bibfield  {title} {\bibinfo {title} {Harnessing nonlinearity: Predicting
  chaotic systems and saving energy in wireless communication},\ }\href@noop {}
  {\bibfield  {journal} {\bibinfo  {journal} {Science}\ }\textbf {\bibinfo
  {volume} {304}},\ \bibinfo {pages} {78} (\bibinfo {year} {2004})}\BibitemShut
  {NoStop}%
\bibitem [{\citenamefont {Pathak}\ \emph {et~al.}(2018)\citenamefont {Pathak},
  \citenamefont {Hunt}, \citenamefont {Girvan}, \citenamefont {Lu},\ and\
  \citenamefont {Ott}}]{Pathak2018}%
  \BibitemOpen
  \bibfield  {author} {\bibinfo {author} {\bibfnamefont {J.}~\bibnamefont
  {Pathak}}, \bibinfo {author} {\bibfnamefont {B.}~\bibnamefont {Hunt}},
  \bibinfo {author} {\bibfnamefont {M.}~\bibnamefont {Girvan}}, \bibinfo
  {author} {\bibfnamefont {Z.}~\bibnamefont {Lu}},\ and\ \bibinfo {author}
  {\bibfnamefont {E.}~\bibnamefont {Ott}},\ }\bibfield  {title} {\bibinfo
  {title} {Model-free prediction of large spatiotemporally chaotic systems from
  data: A reservoir computing approach},\ }\href@noop {} {\bibfield  {journal}
  {\bibinfo  {journal} {Phys. Rev. Lett.}\ }\textbf {\bibinfo {volume} {120}},\
  \bibinfo {pages} {024102} (\bibinfo {year} {2018})}\BibitemShut {NoStop}%
\bibitem [{\citenamefont {Carroll}\ and\ \citenamefont
  {Pecora}(2019)}]{Carroll2019}%
  \BibitemOpen
  \bibfield  {author} {\bibinfo {author} {\bibfnamefont {T.~L.}\ \bibnamefont
  {Carroll}}\ and\ \bibinfo {author} {\bibfnamefont {L.~M.}\ \bibnamefont
  {Pecora}},\ }\bibfield  {title} {\bibinfo {title} {Network structure effects
  in reservoir computers},\ }\href@noop {} {\bibfield  {journal} {\bibinfo
  {journal} {Chaos: An Interdisciplinary Journal of Nonlinear Science}\
  }\textbf {\bibinfo {volume} {29}},\ \bibinfo {pages} {083130} (\bibinfo
  {year} {2019})}\BibitemShut {NoStop}%
\bibitem [{\citenamefont {Press}\ \emph {et~al.}(2007)\citenamefont {Press},
  \citenamefont {Teukolsky}, \citenamefont {Vetterling},\ and\ \citenamefont
  {Flannery}}]{NR}%
  \BibitemOpen
  \bibfield  {author} {\bibinfo {author} {\bibfnamefont {W.~H.}\ \bibnamefont
  {Press}}, \bibinfo {author} {\bibfnamefont {S.~A.}\ \bibnamefont
  {Teukolsky}}, \bibinfo {author} {\bibfnamefont {W.~T.}\ \bibnamefont
  {Vetterling}},\ and\ \bibinfo {author} {\bibfnamefont {B.~P.}\ \bibnamefont
  {Flannery}},\ }\href@noop {} {\emph {\bibinfo {title} {Numerical Recipes 3rd
  Edition: The Art of Scientific Computing}}},\ \bibinfo {edition} {3rd}\ ed.\
  (\bibinfo  {publisher} {Cambridge University Press},\ \bibinfo {address} {New
  York, NY, USA},\ \bibinfo {year} {2007})\BibitemShut {NoStop}%
\bibitem [{Sup()}]{Supplement}%
  \BibitemOpen
  \href@noop {} {}\bibinfo {note} {Consult Supplementary Material at [link
  TBA].}\BibitemShut {Stop}%
\bibitem [{\citenamefont {Strogatz}(2015)}]{Strogatz}%
  \BibitemOpen
  \bibfield  {author} {\bibinfo {author} {\bibfnamefont {S.~H.}\ \bibnamefont
  {Strogatz}},\ }\href
  {https://search.library.wisc.edu/catalog/9910223127702121} {\emph {\bibinfo
  {title} {Nonlinear Dynamics and Chaos: with Applications to Physics, Biology,
  Chemistry, and Engineering}}}\ (\bibinfo  {publisher} {Second edition.
  Westview Press},\ \bibinfo {year} {2015})\BibitemShut {NoStop}%
\bibitem [{\citenamefont {Melnikov}\ and\ \citenamefont
  {Shevchenko}(2004)}]{Melnikov2004}%
  \BibitemOpen
  \bibfield  {author} {\bibinfo {author} {\bibfnamefont {A.~V.}\ \bibnamefont
  {Melnikov}}\ and\ \bibinfo {author} {\bibfnamefont {I.~I.}\ \bibnamefont
  {Shevchenko}},\ }\bibfield  {title} {\bibinfo {title} {The maximum lyapunov
  exponent of the chaotic motion in the h{\'e}non-heiles problem},\ }in\
  \href@noop {} {\emph {\bibinfo {booktitle} {Order and Chaos in Stellar and
  Planetary Systems}}},\ Vol.\ \bibinfo {volume} {316}\ (\bibinfo {year}
  {2004})\ p.~\bibinfo {pages} {34}\BibitemShut {NoStop}%
\bibitem [{\citenamefont {Skokos}(2001)}]{Skokos_2001}%
  \BibitemOpen
  \bibfield  {author} {\bibinfo {author} {\bibfnamefont {C.}~\bibnamefont
  {Skokos}},\ }\bibfield  {title} {\bibinfo {title} {Alignment indices: a new,
  simple method for determining the ordered or chaotic nature of orbits},\
  }\href {https://doi.org/10.1088/0305-4470/34/47/309} {\bibfield  {journal}
  {\bibinfo  {journal} {Journal of Physics A: Mathematical and General}\
  }\textbf {\bibinfo {volume} {34}},\ \bibinfo {pages} {10029} (\bibinfo {year}
  {2001})}\BibitemShut {NoStop}%
\bibitem [{\citenamefont {Wang}\ \emph {et~al.}(2019)\citenamefont {Wang},
  \citenamefont {Zhai},\ and\ \citenamefont {You}}]{wang2019emergent}%
  \BibitemOpen
  \bibfield  {author} {\bibinfo {author} {\bibfnamefont {C.}~\bibnamefont
  {Wang}}, \bibinfo {author} {\bibfnamefont {H.}~\bibnamefont {Zhai}},\ and\
  \bibinfo {author} {\bibfnamefont {Y.-Z.}\ \bibnamefont {You}},\ }\bibfield
  {title} {\bibinfo {title} {Emergent quantum mechanics in an introspective
  machine learning architecture},\ }\href@noop {} {\bibfield  {journal}
  {\bibinfo  {journal} {arXiv:1901.11103}\ } (\bibinfo {year}
  {2019})}\BibitemShut {NoStop}%
\bibitem [{\citenamefont {Schmidt}\ and\ \citenamefont
  {Lipson}(2009)}]{Schmidt81}%
  \BibitemOpen
  \bibfield  {author} {\bibinfo {author} {\bibfnamefont {M.}~\bibnamefont
  {Schmidt}}\ and\ \bibinfo {author} {\bibfnamefont {H.}~\bibnamefont
  {Lipson}},\ }\bibfield  {title} {\bibinfo {title} {Distilling free-form
  natural laws from experimental data},\ }\href
  {https://doi.org/10.1126/science.1165893} {\bibfield  {journal} {\bibinfo
  {journal} {Science}\ }\textbf {\bibinfo {volume} {324}},\ \bibinfo {pages}
  {81} (\bibinfo {year} {2009})}\BibitemShut {NoStop}%
\bibitem [{\citenamefont {Bondesan}\ and\ \citenamefont
  {Lamacraft}(2019)}]{Bondesan2019}%
  \BibitemOpen
  \bibfield  {author} {\bibinfo {author} {\bibfnamefont {R.}~\bibnamefont
  {Bondesan}}\ and\ \bibinfo {author} {\bibfnamefont {A.}~\bibnamefont
  {Lamacraft}},\ }\bibfield  {title} {\bibinfo {title} {Learning symmetries of
  classical integrable systems},\ }\href@noop {} {\bibfield  {journal}
  {\bibinfo  {journal} {ArXiv:1906.04645}\ } (\bibinfo {year}
  {2019})}\BibitemShut {NoStop}%
\end{thebibliography}%

\end{document}